\documentclass[doublecol]{epl2} 

\usepackage{amsmath,amssymb}

\def\CQG{Class. Quantum Grav.}

\def\PRD{Phys. Rev. D}
\def\GRG{Gen. Rel. Grav.}

\def\JMP{J. Math. Phys.}
\def\CMP{Commun. Math. Phys.}

\def\PRL{Phys. Rev. Lett.}

\def\PREP{Phys. Rep.}

\def\PLA{Phys. Lett. A}

\def\RNC{Riv. Nuovo Cimento}

\newcommand{\boldm}[1]{\mbox{\boldmath $#1$}}
\def\espaitemps{({\cal V},g)}
\def\varietat{{\cal V}}

\def\be{\begin{equation}}
\def\ee{\end{equation}}
\def\bea{\begin{eqnarray}}
\def\eea{\end{eqnarray}}
\def\bean{\begin{eqnarray*}}
\def\eean{\end{eqnarray*}}

\newtheorem{prop}{Proposition}
\newtheorem{result}{Result}

\title{A Reformulation of the Hoop Conjecture}
\shorttitle{Reformulation of the Hoop Conjecture} 

\author{Jos\'e M. M. Senovilla\inst{1}}
\shortauthor{J.M.M. Senovilla}

\institute{                    
  \inst{1} F\'{\i}sica Te\'orica, Universidad del Pa\'{\i}s Vasco,
Apartado 644, 48080 Bilbao, Spain 
}

\abstract{A reformulation of the Hoop Conjecture based on the concept of {\em trapped circle} is presented. The problems of severe compactness in every spatial direction, and of how to superpose the hoops with the surface of the black hole, are resolved.
A new conjecture concerning ``peeling" properties of dynamical/trapping horizons is propounded. A novel geometric Hoop inequality 
is put forward.
The possibility of carrying over the results to arbitrary dimension is discussed.
}

\begin{document}

\maketitle

Among the various ---yet unproven--- beliefs concerning the physics of black holes, a very popular one is the Hoop Conjecture formulated, in a deliberately vague manner, as \cite{T,MTW}: ``black holes with horizons form when, and only when, a mass $M$ gets compacted into a region whose circumference in every direction is  $C \lesssim 4\pi GM/c^2$."

The impreciseness of this statement was 
remarked at its birth \cite{T,MTW}, and has been discussed many times, e.g. \cite{R,F,Wald1}, for many physical and mathematical difficulties arise. The main problems can be summarized as: 

\noindent
1. In practice, it is 
impossible to determine the existence of an event horizon, 
a global concept accessible only to omniscient observers. A remedy 
has been the use of alternative, {\em local}, definitions of horizons, mainly apparent horizons \cite{HE,KH}, based on the concept of closed trapped surface first introduced in \cite{P}. Nowadays, the newer notions of dynamical/trapping horizons \cite{AK,Hay} seem to be a reasonable alternative. One therefore deals with the 
more fundamental notion of (a world tube of) closed marginally trapped surfaces. I will follow this strategy here.

\noindent
2. The use of the symbol $\lesssim$ 
showing the uncertainty
about the 
numeric constants to be used  
--- several possibilities have appeared in the literature, e.g. \cite{B,B1}. 
It has been even argued that different values should be used for the ``when" and the ``only when" parts of the conjecture \cite{F}.

\noindent
3. In general, the ``circumference" of a bounded body is not defined. In fact, one can pass {\em arbitrarily small} hoops around {\em any} concentration of matter by letting appropriate portions of the hoop move relativistically (close to a null curve) 
\cite{Wald1}. A proposed solution in this case is to use a specific slice, usually a particular type of Cauchy hypersurface, e.g. \cite{NST,ST,M,F,CNNS}. See in this respect, however, the illustrative example in \cite{WI}, also \cite{PTW}.

\noindent
4. Another major problem is the definition of mass $M$. Gravitational energy is non-localizable and
one has to take into account gravitational radiation. More problematic, even after fixing a slice, is the idea of ``mass encircled by the hoop",
without solution already in 
Newtonian physics. The objection here is that
there is no notion of region encircled by a hoop (think of a disk or a hemisphere with identical circle as boundary)

Yet, despite all difficulties, the Hoop Conjecture has been 
successful. It was settled in spherical symmetry \cite{BMO,BMO1,BMO2}, and discussed in some special non-spherical cases \cite{M}. Many numerical and/or analytical idealized examples \cite{BIL,BM,B,B1,CGS,Ch,CM,CNNS,F,INSH,NST,PTW,R,ST,Tod,W,W1,YNT}  have given it robust support. These works (and their references) have sometimes 
simultaneously 
considered the cosmic censorship hypothesis \cite{P1,Wald1,L}, sending conflicting signals, but I will not treat this here.
Also, a mathematical 
result was found in \cite{SY} where an upper bound for a ``radius" of bounded matter was linked to lower bounds for its mass {\em density}. Unfortunately, the size measure in \cite{SY} seems to be inadequate, and their criteria rarely met, to describe the extent of black holes ---see examples in \cite{BMO1}.  

Thus, it would be interesting to have a precise formulation, mathematically sound and physically falsifiable, 
while keeping the spirit of the original conjecture. Its main underlying idea
is that black holes must be localized, so that  
matter-energy must be {\em severely compacted in all spatial directions} if horizons are to be formed. But, what is the exact meaning of ``severe compactness"? Can 
this crucial physical idea be formulated in a mathematically correct way? 
For instance, there is no compelling reason to believe that very elongated matter (such as a long cylinder with two hemispherical caps) cannot collapse to eventually form a black hole {\em whenever} the height of the cylinder is small enough ---compared to the enclosed mass--- to produce that ``severe concentration".
In a way, the concept of closed trapped surface \cite{P,HE} captures 
the idea of severe concentration without speaking of mass,
leading to a ``compactness conjecture" \cite{Sei,M}. 
This is related to isoperimetric inequalities \cite{G,P2}. Other type of 
inequalities, guaranteeing the existence of {\em averaged} closed trapped surfaces, were found in \cite{M,MK} in some special situations. 

Actually, by means of the mean curvature vector \cite{Kr,O,MS},
trapped submanifolds of {\em arbitrary} co-dimension
can be defined rigorously \cite{MS}.
This is the new ingredient I want to bring in, because the concept of {\em trapped hoop} can thus be used without 
speaking of its length or its encircled mass. Thus, one can pass the hoops through the designated region and check whether or not they are trapped {\em when instantaneously superposed at relative rest} with the limiting surface---avoiding also any slicing. 

This concept of trapped hoop allows for a sensible re-formulation of the conjecture. 
More importantly, some reformulated statements can be proven if one is willing to enforce stringent conditions on the closed trapped surfaces.
The required stricter types of trapped surface are known as 
{\em totally, (very) strongly, or (very) lightly} trapped surface \cite{S4}, and are characterized by some properties of their future null second fundamental forms.\footnote{Totally trapped surfaces were called strongly trapped in \cite{Kup}. For strongly marginally trapped surfaces the two nomenclatures agree.}

Let $\espaitemps$ be a $4$-dimensional spacetime
with metric tensor $g$ of signature $(-,+,+,+)$. Consider an imbedded connected spacelike surface $S$ \cite{O,HE}, so that its 
first fundamental form $\boldm{\gamma}$ is positive-definite. 
Then, at any $x\in S$ the tangent space has the orthogonal decomposition
$
T_{x}\varietat =T_{x}S\oplus T_{x}S^{\perp}
$
into {\em tangent} and {\em normal} parts. Using this, the {\em shape tensor} of $S$ in $\varietat$ is defined as
$$
\vec{\boldm{K}} : \, \mathfrak{X}(S) \times \mathfrak{X}(S) \longrightarrow \mathfrak{X}(S)^{\perp}, \hspace{3mm}
-\vec{\boldm{K}}(\vec X,\vec Y)\equiv \left(\nabla_{\vec X}\, \vec Y\right)^{\perp}
$$ 
where $\vec X,\vec Y\in \mathfrak{X}(S)$ are smooth vector fields tangent to $S$.

Given any normal direction $\vec n\in \mathfrak{X}(S)^{\perp}$, the {\em second fundamental form} $\boldm{K}_{\vec n}$ of $S$ in $\espaitemps$ relative to $\vec n$ is the 2-covariant symmetric tensor field on $S$ defined by 
$$
\boldm{K}_{\vec n}(\vec X,\vec Y)\equiv g\left(\vec n,\vec{\boldm{K}}(\vec X,\vec Y)\right),
\hspace{1cm} \forall \vec X,\vec Y\in \mathfrak{X}(S)\, .
$$
$S$ has two {\em independent} normal vector fields, and thus
they can be chosen to be null and future-pointing. Call them $\vec{\ell}, \vec k \in \mathfrak{X}(S)^{\perp}$, and add the normalization condition $g(\vec{\ell},\vec{k})=-1$. 
There remains the freedom
\begin{equation}
\vec{\ell} \longrightarrow \vec{\ell}'=\sigma^2 \vec{\ell}, \hspace{1cm}
\vec{k} \longrightarrow \vec{k}'=\sigma^{-2} \vec{k} \label{free}
\end{equation}
where $\sigma^2>0$ is a function 
on $S$. The shape tensor decomposes as
$\vec{\boldm{K}}=-\boldm{K}_{\vec k}\,\, \vec\ell -\boldm{K}_{\vec\ell}\,\, \vec k \, 
$ where
$\boldm{K}_{\vec k}, \boldm{K}_{\vec\ell}$ are the future null second fundamental forms.

The {\em mean curvature vector} $\vec H$ of $S$ in $\espaitemps$ is an averaged version of the shape tensor defined by
$
\vec H \equiv \mbox{tr}\,  \vec{\boldm{K}}$,
$\vec H \in \mathfrak{X}(S)^{\perp}$,
where the trace tr is taken with respect to $\boldm{\gamma}$. 
The scalars $g(\vec H, \vec n)$= tr$\boldm{K}_{\vec n}$ are termed ``expansion 
along $\vec n$'' of $S$. In particular, $\theta_{\vec k}\equiv \mbox{tr}\, \boldm{K}_{\vec k}$ and 
$\theta_{\vec\ell}\equiv \mbox{tr}\, \boldm{K}_{\vec\ell}$ are called the {\em future null expansions}
and $\vec H$ becomes
\be
\vec{H}= -\theta_{\vec k}\,\, \vec\ell - \theta_{\vec\ell}\,\, \vec k \, .\label{mean}
\ee
$S$ is said to be (semi-) contracting at $x\in S$ if $\vec H|_x$ is (non-zero null) timelike and future-pointing \cite{S4}. If these properties are kept on the whole $S$ then $S$ is called (marginally) future trapped. 
Obviously, $S$ is future trapped if and only if $\theta_{\vec k}<0$ and $\theta_{\vec\ell}<0$,
and marginally future trapped if one of them vanishes and the other is negative, on $S$.
Physically, contracting points have local neighborhoods with initially decreasing area along {\em any} future direction. 

One can also define the vector density \cite{S4}
\be
\vec G \equiv -(\det \boldm{K}_{\vec k})\, \vec \ell - (\det \boldm{K}_{\vec \ell})\, \vec k,
\hspace{1cm}  \vec G \in \mathfrak{X}(S)^{\perp} \, .
\label{det} 
\ee
(\ref{mean}) is invariant under
(\ref{free}), but {\em not} (\ref{det}). However, the causal orientation of $\vec G$ 
is invariant under (\ref{free}). 
The combination of $\vec H$ and $\vec G$ provides a refined classification of trapped surfaces \cite{S4}, in particular a future trapped $S$ is called:

\noindent
--- {\em totally future trapped} if $\vec G$ is past-pointing timelike on $S$; equivalently, both $\det \boldm{K}_{\vec k}$ and $\det \boldm{K}_{\vec \ell}$ are positive. Thus $\boldm{K}_{\vec k}$ and $\boldm{K}_{\vec \ell}$ are negative definite matrices.

\noindent
--- {\em strongly future trapped} if $\vec G$ is past-pointing on $S$; equivalently, $\det \boldm{K}_{\vec k}$ and $\det \boldm{K}_{\vec \ell}$ are non-negative, so that $\boldm{K}_{\vec k}$ and $\boldm{K}_{\vec \ell}$ are negative semi-definite. 
If in addition $\vec G$ is non-zero on $S$, then $S$ is called {\em very strongly future trapped}. In this case, $\det \boldm{K}_{\vec k}$ and $\det \boldm{K}_{\vec \ell}$ do not vanish simultaneously.

Similarly, a marginally future-trapped surface (so that $\vec H\neq \vec 0$ is null and future pointing) is said to be:

\noindent
--- {\em totally marginally future-trapped} if $\vec G$ is past null (necessarily) collinear with $\vec H$; equivalently, one of $\det \boldm{K}_{\vec k}$ or $\det \boldm{K}_{\vec \ell}$ vanishes and the other is positive. Hence, one of $\boldm{K}_{\vec k}$ or $\boldm{K}_{\vec \ell}$ vanishes and the other is negative definite. 

\noindent
--- {\em very strongly marginally future-trapped} if $\vec G$ is past causal (necessarily collinear with $\vec H$): one of $\det \boldm{K}_{\vec k}$ or $\det \boldm{K}_{\vec \ell}$ vanishes, the other is non-negative. Hence, one of $\boldm{K}_{\vec k}$ or $\boldm{K}_{\vec \ell}$ vanishes, the other is negative semi-definite.

Strongly or higher (marginally) trapped surfaces are such that
the norms of all their tangent vectors are non-increasing along {\em any} orthogonal future direction. Some examples of these surfaces in well-known black hole solutions are mentioned below ---after Result \ref{onlyif1}---, see also \cite{S4}.

Now, let ${\cal C}$ be a differentiable curve in $S$. Regarded as a curve in the whole $\varietat$, ${\cal C}\subset S$ is obviously a spacelike curve. Let $\vec v\in \mathfrak{X}({\cal C})\subset \mathfrak{X}(S)$ be the {\em unit} tangent vector to ${\cal C}$, then the acceleration (or curvature) of ${\cal C}$ in $\varietat$ is
\be
\nabla_{\vec v}\vec v=
-\vec{\kappa}(\vec v)-\vec{\boldm{K}}(\vec v,\vec v)\equiv -\vec{\varkappa}(\vec v).\label{acc}
\ee
Here $\vec{\kappa}(\vec v)\in \mathfrak{X}({\cal C})^{\perp}\cap \mathfrak{X}(S)$ is the acceleration of ${\cal C}$ as a curve in $(S,\boldm{\gamma})$, and $\vec{\varkappa}(\vec v)\in \mathfrak{X}({\cal C})^{\perp}$. $\vec{\kappa}(\vec v)$ and $\vec{\varkappa}(\vec v)$ are actually the mean curvature vectors 
of ${\cal C}$ in $S$ and $\varietat$, respectively. 
${\cal C}$ is therefore (marginally) future trapped whenever $\vec{\varkappa}(\vec v)$ is (null) timelike and future pointing \cite{MS}. 

Take now two curves ${\cal C}_1$ and ${\cal C}_2$ in $S$, and assume that they meet {\em orthogonally} at a point $x\in {\cal C}_1\cap {\cal C}_2\subset S$. Using (\ref{acc}) for both curves 
we obtain
\be
\vec H|_x=\left.\left(\vec{\varkappa}(\vec v_1)-\vec{\kappa}(\vec v_1)\right)\right|_x+\left.\left(\vec{\varkappa}(\vec v_2)-\vec{\kappa}(\vec v_2)\right)\right|_x \, .\label{H1}
\ee
Since
$\vec H$ is orthogonal to both 
$\vec{\kappa}(\vec v_1)\propto \vec v_2$
and $\vec{\kappa}(\vec v_2)\propto \vec v_1$,
$\vec H$ is future pointing whenever so are $\vec{\varkappa}(\vec v_1)$ and $\vec{\varkappa}(\vec v_2)$.
However, this is too strong a requirement, as it is clear from (\ref{H1}): 
it is sufficient that
$\vec{\varkappa}(\vec v_1)-\vec{\kappa}(\vec v_1)$ and $\vec{\varkappa}(\vec v_2)-\vec{\kappa}(\vec v_2)$ be future-pointing 
to get the same conclusion. Noting that
$\vec{\varkappa}(\vec v)-\vec{\kappa}(\vec v)= \vec{\boldm{K}}(\vec v,\vec v) \in \mathfrak{X}(S)^{\perp}$ is the projection
{\em orthogonal} to $S$ of the mean curvature vector of ${\cal C}$ in $\varietat$, the following definition is in order: any ${\cal C}\subset S$ with a future-pointing (null) timelike $\vec{\varkappa}(\vec v)-\vec{\kappa}(\vec v)$ will be called 
{\em $S$-outwardly (marginally) future trapped}.  

Thus, the following elementary result holds:

\noindent
{\em If two $S$-outwardly future trapped curves in $S$ meet orthogonally at $x\in S$, then $S$ is contracting at $x$. While if two $S$-outwardly marginally future trapped curves in $S$ are orthogonal at $x\in S$, then $S$ is either semi-contracting or contracting at $x$ depending on whether $\vec{\varkappa}(\vec v_1)-\vec{\kappa}(\vec v_1)$ and $\vec{\varkappa}(\vec v_2)-\vec{\kappa}(\vec v_2)$ are collinear at $x$ or not. } (If they are collinear, they are called {\em consistently} marginally trapped.)

{\bf Remarks:} (1) The first conclusion is stronger than stated: not only $S$ is contracting ($\theta_{\vec k}|_x<0$ and $\theta_{\vec\ell}|_x<0$), also $\boldm{K}_{\vec k}(\vec v_i,\vec v_i)|_x<0$ and 
$\boldm{K}_{\vec \ell}(\vec v_i,\vec v_i)|_x<0$ for $i=1,2$. This implies that $\vec{\boldm{K}}(\vec v_i,\vec v_i)|_x$ are future timelike for both orthogonal directions $\vec v_i$. This can happen only for some types of contracting points 
(Appendix in \cite{S4}).
(2) If {\em every} curve 
tangent to $S$ at $x\in S$ is $S$-outwardly future trapped, then $\boldm{K}_{\vec k}|_x$ and $\boldm{K}_{\vec \ell}|_x$ are negative definite matrices.

The previous elementary result is local and independent of the topology 
of $S$. It also applies to planes ($\mathbb{R}^2$) or cylinders ($S^1\times \mathbb{R}$), so any hoop idea is yet missing. Recall that there are non-compact trapped surfaces with those topologies
in flat space-time (e.g.\ Example 4.1 in \cite{S}, p.776), and thus they cannot be related to horizons or black holes. 

To 
link trapped circles with the Hoop Conjecture 
one has to deal with compact $S$ and use closed loops ${\cal C}$ 
that ``go round $S$" ---not just local loops near $x$. One possibility is to use, for each $x\in S$, {\em congruences} of loops emanating from $x$ which cover the entire $S$ \cite{F}, then selecting their members not contained in any convex neighborhood (in $S$). The required hoops will be those with greater lengths on each possible family.
Another possibility would be to employ closed geodesics in $S$. Periodic geodesics 
intuitively seem to do the job (e.g. in the standard round sphere they are the great circles\footnote{
However, in the standard torus $S^1\times S^1$ all its meridians are closed geodesics but, letting aside the inner and outer equators, their corresponding orthogonal closed geodesics go ``up and down" crossing the outer equator. To use them as good models of a hoop is debatable.
Nonetheless, another simpler possibility in this case is to use the topological foliation defined by a set of meridians and their everywhere orthogonal circles. 
}).
It is interesting to remark that
$\vec{\kappa}(\vec v)=\vec 0$ for geodesics, hence $S$-{\em outwardly} trapped geodesics are simply trapped.
Closed geodesics 
in relation with the Hoop Conjecture were 
used in \cite{CNNS,Ch}.
Unfortunately, their existence 
for arbitrary 
$x\in S$ is not guaranteed for the important case of a topological sphere.\footnote{ 
For general closed surfaces $S$ with any smooth Riemannian metric, 
any non-trivial free homotopy class has a closed geodesic, from which 
there are in fact infinitely many closed geodesics for non-$S^2$ surfaces \cite{Kl}. The problematic case is that of surfaces of spherical topology $S^2$ ---the most important case due to the topological censorship results \cite{H,HE,CG,HOY}. 
It is now known \cite{Ba,Fr,Hi} that {\em independently} of the Riemannian metric equipping $S^2$, 
there always exist an {\em infinite} number of closed geodesics. Unfortunately, one does not know how many of them pass through a particular given point.} 

I will use the term {\em hoop} for the appropriate closed loops ``going round" $S$, and I take their abundance for granted. These are not rigid hoops, rather they fit the surface, adapting themselves in order to conform with the shape of $S$. 
Applying the previous results to any closed 
spacelike surface $S$, the following preliminary result holds
\begin{prop}
\label{if}
If two (consistently marginally) $S$-outwardly future-trapped orthogonal hoops pass through every $x\in S$, then $S$ is (marginally) future trapped. 

If all hoops are $S$-outwardly (marginally)
future trapped, then $S$ is totally (marginally) future trapped.
\end{prop}

The converse of the last statement in this proposition also holds. 
For any two unit orthogonal vectors $\vec v_1,\vec v_2\in T_xS$, two curves 
tangent to $\vec v_1$ and $\vec v_2$ will have future-pointing $\vec{\varkappa}(\vec v_1)-\vec{\kappa}(\vec v_1)|_x$ and $\vec{\varkappa}(\vec v_2)-\vec{\kappa}(\vec v_2)|_x$ if
\be
\boldm{K}_{\vec k}(\vec v_i,\vec v_i)\leq 0\, , \hspace{3mm}
\boldm{K}_{\vec\ell}(\vec v_i,\vec v_i)\leq 0 \hspace{3mm} \mbox{for $i=1,2$}.
\label{cond}
\ee
For totally (marginally) future trapped surfaces, these conditions are trivially 
met for {\em all} vectors $\vec v\in T_xS$ ergo:
\begin{prop}
\label{onlyif2}
If $S$ is a closed totally (marginally) future-trapped surface, 
all possible hoops through all $x\in S$ are $S$-outwardly (consistently marginally) future trapped.
\end{prop}

Actually, 
the ``total" condition can be relaxed. 
For (very) strongly future-trapped surfaces, 
$\vec{\boldm{K}}(\vec v,\vec v)$ is timelike and future pointing for all $\vec v\in T_xS$ except for, {\em at most}, (one) two directions at each $x\in S$ \cite{S4}. Analogously, if $S$ is very strongly marginally future trapped, then $\vec{\boldm{K}}(\vec v,\vec v)$ are consistently null and future pointing for all 
$\vec v\in T_xS$ except for, {\em at most}, one direction at each $x\in S$. In both situations it is obvious that almost all possible hoops are $S$-outwardly (marginally) future-trapped.
Observe that if a hoop tends to be tangent to any of the exceptional tangent vectors,
it is enough to slightly modify the hoop, keeping its essential properties, to avoid
this tangency. Thus
\begin{prop}
\label{onlyif3}
If $S$ is a closed strongly future-trapped surface, almost every pair of mutually orthogonal 
hoops through every $x\in S$ is $S$-outwardly future trapped.

If $S$ is a closed very strongly marginally future-trapped surface, almost every pair of mutually orthogonal hoops through every $x\in S$ is $S$-outwardly consistently marginally future trapped.
\end{prop}

A strict converse of the first statement in Proposition \ref{if} encounters some technical difficulties, as conditions (\ref{cond}) may not hold for general future-trapped surfaces. The problem is that a future-pointing $\vec H$ does not imply, in general, that both $\vec{\varkappa}(\vec v_1)-\vec{\kappa}(\vec v_1)$ and $\vec{\varkappa}(\vec v_2)-\vec{\kappa}(\vec v_2)$ are future-pointing.
It is easy to prove, though, that (\ref{cond})
do hold {\em at least} for one of the two null second fundamental forms by choosing the $\vec v_i$ appropriately \cite{S4}: diagonalize (say) $\boldm{K}_{\vec k}|_x$, so that if $\vec e_1,\vec e_2\in T_xS$ are its unit eigenvectors and $\lambda_1,\lambda_2$ the corresponding eigenvalues, then
$\theta_{\vec k}|_x=\lambda_1+\lambda_2\leq 0$. 
Defining two new orthogonal unit vectors by
$$
\left(\begin{array}{c}\vec v_1\\
\vec v_2\end{array}\right) =\left(\begin{array}{cc}\cos\beta & -\sin\beta\\
\sin\beta & \cos\beta
\end{array}\right)
\left(\begin{array}{c}\vec e_1\\
\vec e_2\end{array}\right) 
$$
one gets
$
-\boldm{K}_{\vec k}(\vec v_1,\vec v_1)=-\lambda_1\, \cos^2\beta -\lambda_2\, \sin^2\beta$,
$-\boldm{K}_{\vec k}(\vec v_2,\vec v_2)=-\lambda_2\, \cos^2\beta -\lambda_1\, \sin^2\beta$
which will be both non-negative 
for  {\em at least} the value $\beta =\pi/4$. 
The same can be proven for $\boldm{K}_{\vec\ell}$, leading to {\em another} set of pairs of orthonormal vectors $\vec v'_1,\vec v'_2\in T_xS$. 
But, if the sets of allowed directions $\vec v_1,\vec v_2$ for $\boldm{K}_{\vec k}$ and $\vec v'_1,\vec v'_2$ for $\boldm{K}_{\vec\ell}$ do not have pairs in common, then conditions (\ref{cond}) will not hold simultaneously, and therefore there will be $x\in S$ such that {\em no} pair of $S$-outwardly future-trapped curves is orthogonal at $x$.

To overcome these impediments one needs the existence of very {\em few} directions $\vec v$ for which $\vec{\boldm{K}}(\vec v,\vec v)$ is not future pointing. 
A good possibility to achieve this consists in restricting oneself to {\em very lightly} ({\em strongly} marginally) future trapped surfaces \cite{S4}, characterized by having {\em one} of the null second fundamental forms negative semi-definite (vanishing). If $\boldm{K}_{\vec k}$ is {\em not} the semi-definite (vanishing) matrix, it all depends on the magnitude at each point of the angle 
$$
\arcsin\sqrt{\lambda_1/(\lambda_1-\lambda_2)}
$$
(or the corresponding angle for $\boldm{K}_{\vec \ell}$ in the other case) where $\lambda_1$ is the negative eigenvalue and $|\lambda_1|> \lambda_2> 0$. Let us call this value the {\em critical angle}.
The larger this angle, the better. Thus, as a complement to Propositions \ref{onlyif2} and \ref{onlyif3} there is a stronger result of the following kind:
\begin{result}
\label{onlyif1}
If the critical angle of a closed very lightly (strongly marginally) future-trapped surface $S$ is bounded below  
by a value near enough $\pi/2$, then 
two mutually orthogonal  (consistently marginally) $S$-outwardly future-trapped hoops
pass through every $x\in S$.
\end{result}

Propositions \ref{if}, \ref{onlyif2}, \ref{onlyif3} and Result \ref{onlyif1} provide if-and-only-if
formulations of (a) the notion ``severely compacted in every direction"
and (b) how to superpose the hoops avoiding any slicing.
The pairs of mutually orthogonal and $S$-outwardly (marginally) future-trapped hoops build up a ``cage" confining $S$. For strongly or totally 
trapped $S$, they actually add up to a perfectly fitting adjustable dress.

An open question 
is the exact lower bound for the critical angle defining the sharpest 
combination of Result \ref{onlyif1} and the first statement in Proposition \ref{if}. Nevertheless, this might turn out to be
irrelevant 
if we accept that, in realistic collapses, dynamical/trapping horizons in black hole spacetimes will eventually approach the event horizon and, therefore, be foliated by very strongly 
marginally trapped surfaces. This is supported by several 
results: (i) the probable eventual  
mergence of event horizons and hypersurfaces foliated by marginally trapped closed surfaces \cite{BBGV,AG}, 
discussed 
in \cite{I} and 
partly confirmed 
under some circumstances or numerically, e.g. \cite{AK,AG,many,Boo,Wil} and references therein;
(ii) the black hole ``rigidity" \cite{HE,Chrus} and uniqueness theorems,  e.g. \cite{Heu,Rob} and their references, the former stating that event horizons of stationary black holes are Killing horizons, the latter that electrovacuum black holes in equilibrium are of Kerr-Newman type;
(iii) the fact that non-expanding horizons, including isolated and Killing horizons, are foliated by strongly (or higher) marginally trapped surfaces \cite{AK,GJ}; in particular the Kerr-Newman event horizon is foliated by {\em totally} marginally trapped surfaces. Observe that tubes foliated by {\em strictly stable} \cite{AMS} marginally trapped surfaces are completely partitioned into isolated and dynamical horizons \cite{AMS,BF}; 
(iv) the properties of spherically symmetric horizons, necessarily foliated by 
totally marginally future-trapped spheres \cite{S4}.

Taking into account 
that dynamical horizons have a {\em unique} foliation by marginally trapped surfaces \cite{AG}, I would like to put forward a new conjecture, advanced in \cite{S4}, concerning the ``peeling behavior" of these horizons: {\em their
closed marginally future-trapped surfaces 
become increasingly trapped and eventually are very strongly 
future trapped, whenever the physical system is realistic and leads 
to the formation of a black hole in equilibrium.} 

Propositions \ref{if}, \ref{onlyif2}, \ref{onlyif3} and Result \ref{onlyif1} are purely geometrical results, independent of energy conditions or field equations. 
These are however fundamental for singularity theorems \cite{P,HE,HP,S} and for the issues mentioned in the previous two paragraphs. Hence, 
they are needed to 
promote the results in this paper
into a {\em physical statement}, and for the previous conjecture. 
As
the results only 
resolve 
geometrical problems of the Hoop Conjecture,
one main ingredient is still missing: the relation between geometry and matter,
or how to produce a formula relating a {\em length} with the {\em mass} of a black hole once energy conditions have been assumed. This problem can be addressed: an 
appropriate length
can be defined,
in the frame of the reformulation, by considering all possible $S$-outwardly marginally future-trapped hoops in a marginally future-trapped $S$ 
such that the projection orthogonal to $S$ of their mean curvature vector is parallel to
$\vec H$ 
(consistency). For each $x\in S$, take the infimum of the lengths of all such hoops ${\cal H}_x$ from $x$
and denote it
by $D({\cal H}_x)$. Now, simply put by definition\footnote{This is reasonable only for  $S^2$ topology. Otherwise, each non-trivial free homotopy class has a length $C_k$:
the supremum for $x\in S$ of the infima of the lengths of ${\cal H}_x$ in that class. 
The number of $C_k$ depends on the genus of $S$. 
However, finding a geometric measure of enclosed mass is also problematic in these cases.}
$$
C\equiv \sup_{x\in S} D({\cal H}_x)
$$
which is generically smaller than twice the diameter of $S$
(the supremum of distances between pairs of points).

There only remains to find an appropriate measure for the ``mass''.
This can be estimated by means of {\em any} of the  available quasi-local masses
\cite{Sza}. A fundamental requirement for all of them
is that marginally trapped surfaces with $S^2$-topology enclose a quasi-local mass equal to its so-called irreducible mass (\cite{Sza}, sect.\ 4.3.2, 1.7), that is 
\be
M_{irr}=(c^2/4G) \sqrt{\mbox{Area}(S)/\pi}.\label{mass}
\ee
One may be tempted to write now the formula 
\be
C\leq  \sqrt{\pi\, \mbox{Area}(S)}\label{static}
\ee
so that this coincides, via (\ref{mass}), with the original Hoop formulation where the mass is $M_{irr}$. This formula appears to be true in {\em static} cases. Unfortunately, it is a matter of simple calculation to check that 
(\ref{static}) fails for Kerr black holes near the extreme case.
Hence, we need a better estimate of the right-hand side of the inequality (\ref{static}).

In non-static cases the main new, and fundamental, ingredient is rotation, therefore
angular momentum has to be taken into consideration. It is well-known, e.g. \cite{Ha,PT,AK,GJ}, that angular momentum is related to the one-form $\boldm{\theta}\in \Lambda^1 S$ defined as 
$
\boldm{\theta}(\vec X)\equiv g(\vec \ell , \nabla_{\vec X}\vec k) 
$
for every 
$\vec X\in \mathfrak{X}(S)$.
This one-form is a connection on the normal bundle of $S$, and its change under (\ref{free}) is \cite{S4} $\boldm{\theta}'=\boldm{\theta}+d\log \sigma^2\, .$
Thus, its curvature $d\boldm{\theta}\in \Lambda^2 S$ is invariant under (\ref{free}). 
The standard way to define angular momentum is to consider a fixed 
rotational vector field $\vec \xi \in \mathfrak{X}(S)$ and integrating $\boldm{\theta}(\vec\xi)$ on the compact $S$. This, of course, depends on the representative chosen for $\vec\ell$ \cite{AK}, so it is not invariant under (\ref{free}) ---unless $\vec\xi$ is divergence-free.
Moreover, this requires a preferred rotational 
$\vec\xi$ on  $S$. These properties are not desirable for the geometric formulation that I have in mind.

However, the invariant 2-form $d\boldm{\theta}$ can be used for our purposes. As a matter of fact, this 2-form is known to be related to angular momentum {\em and} gravitational radiation, see \cite{Epp}, which is the second new ingredient in non-static situations. First, define its magnitude as usual: 
${\cal R}^2 \equiv (1/2) \boldm{\gamma}(d\boldm{\theta},d\boldm{\theta})=2\, \theta_{[A;B]}\theta^{[A;B]}\geq 0.$
Normalize it with respect to the curvature on $S$ dividing by 
$Q\equiv \sup_{x\in S} K_{S} > 0$
where $K_{S}$ is the Gaussian curvature of $S$. Integrate this over the compact $S$ to obtain the quantity
$$
\mbox{Rot}(S)\equiv \frac{1}{Q}\int_{S} {\cal R}\, .
$$
With these well-posed definitions for $C$ and Rot$(S)$, a geometric Hoop formula can be explicitly written as
\be
C\leq  \sqrt{\pi \left[\mbox{Area}(S) + b\,   \mbox{Rot}(S)\right]}\label{hoop}
\ee
where $b$ is a positive number. It is feasible in principle that this inequality held for untrapped surfaces (if the angular momentum is very large for a given mass), so that for the ``only when'' part of the conjecture $S$ in (\ref{hoop}) should probably be understood as (approaching) a very strongly marginally future-trapped $S^2$-surface so that the ``peeling off" has taken place and the critical angle is large enough. In order to fix $b$ one can use the standard Kerr black holes. An elementary calculation proves that the above inequality is saturated for the cases of (i) the Schwarzschild black hole (Rot$(S)$=0) {\em and} (ii) the extreme Kerr solution, in both cases exactly for the value
$$
b=2\, .
$$
For this value, the inequality is strict in any other Kerr black hole.
One wonders whether the right-hand side of the inequality (\ref{hoop}) may lead to a new definition of ``mass'', but this is outside the scope of this letter.

Let me finally consider
whether the reformulation can be extended to 
dimensions $n\geq 4$. By carefully dissecting the reasonings used to prove the results
one realizes that higher dimensions 
request families of closed $(n-3)$-dimensional submanifolds, in agreement with 
\cite{IN,YN,BFL}.
The generalization of the basic relation (\ref{H1}) to higher values of $n$ is then straightforward. Take $n-2$ hypersurfaces ${\cal C}_j\subset S$ of a co-dimension 2 embedded submanifold $S$ such that they meet mutually orthogonally at $x\in S$.
Then 
$$
\vec H|_x=\frac{1}{n-3}\sum_{j=1}^{n-2}(\vec{\varkappa}_j-\vec{\kappa}_j)|_x
$$
where $\vec{\varkappa}_j$ and $\vec{\kappa}_j$ are the mean curvature vectors of each ${\cal C}_j$ 
in $\varietat$ and in $S$ respectively. 
However, the election of the ``hoops" ${\cal C}_j$ must be done with care depending on the topology of $S$, see \cite{CG,HOY}. For $S^{n-2}$-topology
one has to choose $(n-3)$-dimensional spheres, but for a black ring \cite{ER}
with topology $S^1\times S^{n-3}$ the choice is one hoop with topology $S^{n-3}$ and $n-3$ hoops with topology $S^1\times S^{n-4}$. 

Other difficulties may arise due to the peculiar properties of {\em spherical} black holes in higher dimensions. 
For instance, in dimensions 
$n>5$ the horizon may have very different sizes in different directions, 
with arbitrarily large angular momentum for a given mass
resulting in a tiny $(n-2)$-volume for $S$ \cite{EM}.
However, 
this volume scales as $Mr_{+}$, where $M$ is the mass and $r_{+}$ 
is typically of the order of the horizon's {\em small} characteristic length,
ergo an inequality of the type shown in this paper might still be possible for such ultra-spinning black holes.

\acknowledgments
I am indebted to R Emparan, E Malec, M Mars, M S\'anchez and R Vera for many
suggestions. I acknowledge support from grants FIS2004-01626 and GIU06/37.

\end{document}